\documentclass[epj]{svjour}
\usepackage{graphics}
\usepackage{amssymb}
\usepackage{epsfig}
\newcommand{\ga}{\alpha}
\newcommand{\gb}{\beta}

\newcommand{\gC}{\Gamma}

\newcommand{\gl}{\lambda}
\newcommand{\gs}{\sigma}

\newcommand{\gL}{\Lambda}

\newcommand{\be}{\begin{equation}}
\newcommand{\ee}{\end{equation}}
\newcommand{\ba}{\begin{eqnarray}}
\newcommand{\ea}{\end{eqnarray}}
\newcommand{\as}[1]{\langle#1\rangle}

\begin{document}
\title{Shear viscosity of the Quark-Gluon Plasma from a virial expansion}

\author{Stefano Mattiello\inst{1}\thanks{\emph{Present address:} stefano.mattiello@theo.physik.uni-giessen.de} \and Wolfgang Cassing\inst{1}
%
}                     
%
%
\institute{Institute for Theoretical Physics, University of Gie{\ss}en, Germany}
\date{Received: date / Revised version: date}
%
\abstract{
We calculate the shear viscosity $\eta$  in the quark-gluon plasma
(QGP) phase within a virial expansion approach with
particular interest in the ratio of $\eta$ to the entropy density
$s$, i.e. $\eta/s$. The virial expansion approach allows us to
include the interactions between the partons in the deconfined
phase and to evaluate the corrections to a single-particle
partition function. In the latter approach we start with an
effective interaction  with parameters fixed to reproduce
thermodynamical quantities of QCD such as energy and/or entropy
density. We also directly extract the effective coupling $\ga_{\rm
V}$ for the determination of $\eta$. Our numerical results give a
ratio $\eta/s\approx 0.097$ at the critical temperature $T_{\rm c}$,
which is very close to the theoretical bound of $1/(4\pi)$.
Furthermore, for temperatures $T\leq 1.8 T_{\rm c}$ the ratio
$\eta/s$ is in the range of the present experimental estimates
$0.1-0.3$ at RHIC. When combining our results for $\eta/s$ in the
deconfined phase with those from chiral perturbation theory or the
resonance gas model in the confined phase we observe a pronounced minimum of
$\eta/s$ close to the critical temperature $T_{\rm c}$.
} 
\maketitle
\section{Introduction}\label{intro}

The exploration of the phase structure of QCD is a challenging
task for modern theoretical physics. Lattice QCD (lQCD)
calculations for vanishing quark chemical potential $\mu_{\rm q}$ have
shown chiral and deconfinement phase transitions at a critical
temperature $T_{\rm c}$ of about 150 to 200 MeV~\cite{Cheng:2007jq,Aoki:2006br}. New
methods to extend lQCD calculations to finite chemical potentials
$\mu_{\rm q}$ have been developed in the past decade, i.e. a
multi-parameter
reweighting~\cite{Fodor:2001pe,Fodor:2002km,Fodor:2004nz}, a
Taylor expansion around $\mu_{\rm q} \simeq
0$~\cite{Allton:2002zi,Allton:2003vx,Allton:2005gk} and an
imaginary chemical potential method
\cite{D'Elia:2002gd,D'Elia:2004at}; however, their validity is
limited to a region $\mu_{\rm q} \lesssim T$~\cite{Katz:2003up}. At
finite net quark densities presently some modeling of QCD is
needed in order to explore the QCD phase diagram at least on a
qualitative level. Indeed, effective approaches indicate a very
rich phase structure of QCD, i.e. 1) a quark-gluon plasma (QGP),
2) a color superconducting phase, 3) a color-flavor locking phase
and different further combinations (cf. Refs.
~\cite{Alford:1997zt,Alford:1998mk,Klevansky:1992qe,Ratti:2005jh,Cassing:2007nb,Strauss:2009uj}).
The explicit phase structure, however, is model dependent (except
for very high $\mu_{\rm q}$) and not accessible by experiment in the
full $(T, \mu_{\rm q}$) -plane. Experimental information can be
qualitatively obtained from astronomical observations and more
quantitatively from relativistic  heavy-ion collisions. In the
last decade such experiments have been performed at the
Relativistic Heavy-Ion Collider (RHIC) at Brookhaven National
Laboratory (BNL) and even long before - at lower energies - at the
Super-Proton-Synchrotron (SPS) at CERN.

Results from lQCD calculations have early been interpreted as
indicating a weakly interacting system of partons since the
entropy density $s$ and the energy density $\varepsilon$ are close
to the Stefan Boltzmann (SB) limit for a relativistic
noninteracting system, i.e. only  $\approx 10-15\%$ lower.
However, more recent experimental observations at RHIC have
drastically changed the notion about the QGP. In this context, one
of the most intriguing experimental findings is the large elliptic
flow  $v_2$ of hadrons at RHIC
\cite{Ackermann:2000tr,Adcox:2002ms} which is significantly larger
than at SPS energies~\cite{Alt:2003ab,Agakichiev:2003gg}. In
particular the transverse momentum $p_{\rm T}$ dependences of the
elliptic flow $v_2(p_{\rm T})$ at RHIC is close to predictions from
non-dissipative hydrodynamical
simulations~\cite{Kolb:2000fh,Hirano:2001eu,Heinz:2009xj,Teaney:2009qa,Romatschke:2009im} around midrapidity ($| y | \leq 1$).
This result has led to the
BNL announcement about the discovery of the nearly perfect
fluidity of the strongly-coupled quark-gluon plasma (sQGP)
\cite{Gyulassy:2004zy,Shuryak:2004cy} produced at RHIC.
Especially, due to the asymptotic freedom of QCD, the prevailing
idea has been an expectation of large shear viscosities in a
weakly interacting QGP (wQGP) at very high densities and/or
temperatures. Because of the evident failure of these assumptions
at RHIC conditions the novel notion of a strong quark-gluon plasma
(sQGP) has been put forward to characterize the observed strong
coupling properties of the QGP close to (or slightly above)  the
critical temperature $T_{\rm c}$ that keep viscous effects low at RHIC.
Therefore, the inclusion of interactions between partons is
mandatory to consistently describe the QGP in the region close to
$T_{\rm c}$.

In going beyond the standard dynamical quasiparticle picture
~\cite{Cassing:2007nb,Cassing:2007yg,Peshier:2004bv,Peshier:2005pp,Peshier:2004ya}
- incorporating interactions in terms of a width in the spectral
functions of the partons -  we have recently developed a
generalization of the classical virial expansion approach to
calculate the QCD partition function in the partonic phase with an
interaction inspired by lattice calculations~\cite{Mattiello:2009fk}.
We have obtained an Equation-of-State
(EoS) for the partonic QGP that is well in line with recent
three-flavor QCD lattice data~\cite{Cheng:2007jq} for the
pressure, speed of sound and interaction measure at nonzero
temperature and vanishing chemical potential ($\mu_{\rm q} = 0$). Since
in the virial expansion approach all thermodynamic quantities are
based on an explicit parton interaction in form of a potential,
this approach is also the ideal starting point for a consistent
description of dynamical properties of the QGP like the shear
viscosity $\eta$, which is the stationary limit of a nontrivial
correlator~\cite{Karsch:1986cq,Hosoya:1983id}.

In this work we focus on the calculation of the ratio $\eta/s$,
well known as specific viscosity, within the virial expansion approach.
It was shown some years ago for supersymmetric
Yang-Mills (SYM) gauge theory using the Anti de-Sitter
Space/Conformal Field Theory (AdS/CFT) duality
conjecture~\cite{Kovtun:2004de}  that
\begin{equation}
\label{ads} \left( \frac{\eta}{s}\right)_{\mathrm{SYM}} =
\frac{1}{4\pi} \; \; , \end{equation} which is denoted as KSS
bound. This limit is close to the simple bound $\eta/s\geq 1/15$
from the kinetic theory uncertainty
principle~\cite{Danielewicz:1984ww}. Additionally, it has even
been speculated that this bound might hold  for all
substances~\cite{Kovtun:2004de}. Furthermore, it has been found
for atomic and molecular substances that the specific viscosity
exhibits a minimum in the vicinity of the liquid-gas critical
point; this also suggests the possibility of a minimum of $\eta/s$
for QCD at the critical temperature $T_{\rm c}$ which has been
already noted in Refs. \cite{Lacey:2006bc,Csernai:2006zz}.

\section{Kinetic theory}\label{Sec:KinTheo}

We start by recalling results well known from literature.
Here kinetic theory is a convenient framework to start
the investigation of the viscosity coefficient $\eta$~\cite{Lacey:2006bc}.
In an ultrarelativistic quark-gluon plasma,
i.e. where the temperature $T$ is much
larger than the constituent masses $m_i$, the estimate for the
ratio of the shear viscosity  $\eta$ and the entropy density $s$
is~\cite{Danielewicz:1984ww}
\be\label{eqn:Visk-KT}
\frac{\eta}{s}\simeq\frac{1}{s}\frac{4}{15}\sum_{i}(\rho\as{p}\gl)_i\,
\ee where $\rho_i$ is the local density of quanta $i$
transporting on average a momentum $\as{p}_i$ over a
momentum-degradation length (mean-free-path) $\gl_i$.
Following the more detailed kinetic theory
derivations of Ref.~\cite{Reif:1965}, the familiar factor
$\frac{1}{3}$ of elementary non-relativistic kinetic theory has to be
replaced in the ultrarelativistic domain by $4/15$.

Evidently, two aspects play a crucial role in the expression
(\ref{eqn:Visk-KT}) for $\eta/s$.
The first one is the choice of the equation-of-state for the
determination of the entropy density $s$. In general - for such
calculations - an ideal gas EoS is used. This assumption might be
justified for high temperatures $T \gg T_{\rm c}$, where the deviation
of the QGP equation-of-state from the ideal gas limit might be
eventually neglected. On the other hand, close to the critical
temperature $T_{\rm c}$, this approximation should be abandoned due to
large deviations from the SB limit.

The second crucial point is given by partonic dynamics itself: The effects
of the interaction
between the constituents directly determine the mean-free-path
$\lambda$ and its calculation is the aim of several investigations
within different approaches. Following Ref.~\cite{Thoma:1991em}
the mean-free-path is the inverse of the interaction rate for
nearly massless quanta  $\gC_i$, i.e. $\gl_i=1/\gC_i$, which can
be calculated to lowest order in the coupling constant $g$ from
the imaginary part of the quark and gluon selfenergy. Another
possibility is to relate the mean-free-path to the transport cross
section $\gs_{\rm t}$ by $\gl_i=(\rho\gs_{\rm t})^{-1}_i$. This
quantity - in a dense partonic system - may be related to the
transport parameter $\hat q$ governing multiple
scattering~\cite{Majumder:2007zh,Muller:2008bs}
or can be directly modeled in a Debye-screened form~\cite{Zhang:1999rs};
the latter leads to an analytic expression
for $\gs_{\rm t}$~\cite{Molnar:2001ux}. Using this last
formulation and additionally adopting the approximation
$\as{p}\approx 3T$ the specific viscosity becomes \be\label{eqn:Visk-s}
\frac{\eta}{s}\approx\frac{4}{5}\frac{T}{s\gs_{\rm t}} \ . \ee
In particular, assuming that
the elastic gluon scattering matrix element in a dense partonic
medium can be modeled by a Debye screened
interaction~\cite{Zhang:1999rs,Molnar:2001ux} the relevant
transport cross section reads
\ba \label{cross} \sigma_{\rm
t}(\hat s) &\equiv& \int d\sigma_{\rm el} \;\sin^2\theta_{\rm cm}\nonumber\\
&=&\sigma_0 \;
4z(1+z)\left[(2z+1)\ln(1+1/z) - 2\right],
\label{Eq:transport_xs} \ea with the total cross section
$\sigma_0(\hat s) = 9\pi\alpha_{\rm V}^2(\hat s)/2\mu_{\rm scr}^2$.
Here $\ga_{\rm V}=\ga_{\rm V}(T)$ and $\mu_{\rm scr}$ are the effective
temperature dependent coupling constant and the screening mass, respectively, and
$z\equiv\mu_{\rm scr}^2/\hat s$. For simplicity, we
will assume $\sigma_0$ to be energy independent and neglect its
weak logarithmic dependence on $\hat s$ in the relevant energy
range and set $\hat{s}\approx 17 T^2$. We recall that the
$\sin^2\theta_{\rm cm}$ weight arises in the transport cross section
because large-angle scatterings are most effective in momentum
degradation~\cite{Danielewicz:1984ww}. The cross section
(\ref{cross}) is a monotonic function of $\mu_{\rm scr}$, which plays a
crucial role for the results of $\gs_{\rm t}$. It is important to
emphasize that the coupling constant and the Debye mass are not independent
parameters in the  calculation because $\mu_{\rm scr}$ is determined by the
value of the coupling  $\ga_{\rm V}$ itself. In perturbation theory one
explicitly obtains
\be\label{def:screen-mass}
\mu_{\rm scr}^2=4 \pi \alpha_{\rm V} T^2
\ee
in gluon dynamics.

For numerical estimates a specific form for $\alpha_{\rm V}(T)$ has to
be chosen. As a quantitative reference the long-range part of the
strong coupling constant extracted from the free energy of a
quark-antiquark pair in lQCD~\cite{Kaczmarek:2004gv} has been used in the
literature~\cite{Hirano:2005wx}.
We point out that any extraction of a coupling constant $\ga_{\rm V}(T)$ from lQCD is
model dependent and deviations (or agreement) of any
parametrization from (with) the lattice data have to be
considered with care. Accordingly, several parametrizations can be
found in the
literature~\cite{Cassing:2007yg,Cassing:2007nb,Peshier:2004ya,Kaczmarek:2004gv,Hirano:2005wx}.

At this point, we emphasize that in previous works~\cite{Hirano:2005wx} these
two aspects -  equation of state and transport cross section -  are
uncorrelated. Therefore a consistent approach is needed which provides {\em i)}
a realistic equation-of-state - and thus a proper entropy density - as
well as {\em ii)} a transport cross section within the same framework.

\section{The virial expansion approach to QCD}\label{Sec:Model}

As pointed out in the previous Section, a consistent approach requires two
fundamental
ingredients: {\em i)} a calculation of the equation-of-state including the
interactions
between the constituents and {\em ii)} an extraction of the effective coupling
that enters the estimate for the transport cross section.
Both requirements can be achieved within the virial expansion formalism developed in a
previous work~\cite{Mattiello:2009fk}, where a detailed derivation of the
partition function $Z(T,V)$,
of all thermodynamic quantities - such as pressure, entropy density, interaction measure
and sound velocity- and of the EoS of the QGP at vanishing and finite $\mu_
{\rm q}$ has been presented.
We achieve an expansion of $\ln Z$ in powers of the logaritm of the partition
function in the Stefan Boltzmann limit $\ln Z^{(0)}$,
i.e. (cf.~\cite{Mattiello:2009fk})
\begin{equation}
\ln Z\approx \ln Z^{(0)}+\frac{b_2}{2}\left(\ln Z^{(0)}\right)^2
\end{equation}
with the second virial coefficient
\be
b_2=\int_V\!{\rm d}^3 {\bf r}\;\left({\rm e}^{-\gb W_{12}(r)}-1\right).\label{eqn:b_2}
\ee
The pressure is simply obtained as
\begin{equation}\label{eqn:P}
P=T\ln{Z}.
\end{equation}
The other quantities can be calculated from the pressure $P$
or the partition function $Z$ using thermodynamic relations.
In particular, for the entropy density one obtains \cite{Mattiello:2009fk}
\be
s=\frac{\partial P}{\partial T}\label{eqn:S}.
\ee
For an application of this formalism to the QGP a specific choice of the interaction
potential $W_{12}$ has to be made.
Following Ref.~\cite{Mattiello:2009fk} we use an effective quark-quark potential inspired
by a phenomenological
model which includes non-perturbative effects from dimension two gluon condensates (that
reproduce the free energy of quenched QCD very well).
The effective potential between the quarks explicitly reads
\begin{equation} \label{pot}
W_{12}(r,T)=\left(\frac{\pi}{12}\frac{1}{r}+\frac{{\mathcal C}_2}{2N_{\rm c}T}\right){\rm e}^{-M(T)r},
\end{equation}
where ${\mathcal C}_2$ is the non-perturbative dimension two condensate and $M(T)$ a
Debye mass estimated as
\begin{equation}\label{def:Debyemass}
M(T)=\sqrt{N_{\rm c}/3+N_{\rm f}/6}\; gT=\tilde g T,
\end{equation}
where we have neglected any scale dependence in the coupling
constant. By using the potential (\ref{pot}) we assume the
interaction to be the same for quarks and antiquarks and neglect
explicit gluon contributions. The latter are encoded in parametric
form in the interaction (\ref{pot}). In short, we generalize the
Yukawa-liquid model for the QGP investigated in
Refs.~\cite{Thoma:2004sp,Thoma:2005ym} before. For a detailed
explanation of this interaction we refer the reader to
Ref.~\cite{Mattiello:2009fk}. A comparison with three-flavor lQCD
calculations with almost physical masses from
Ref.~\cite{Cheng:2007jq}  shows  that a coupling parameter $\tilde
g=1.30$ allows for a good description of all thermodynamic
quantities in the temperature range from 0.8 $T_{\rm c}$ to 5 $T_{\rm c}$.
A detailed discussion of the validity of the virial expansion
truncated at the second term is given in Ref.~\cite{Mattiello:2009fk}.

In order to calculate a transport cross section with this interaction the coupling $\ga_{\rm V}$
has to be extracted from $V_1$.
Following Ref.~\cite{Kaczmarek:2005ui} we define the coupling in the
so-called ${\rm qq}$-scheme,
\begin{eqnarray}
\alpha_{\rm qq}(r,T)&\equiv&-\frac{12}{\pi}r^2\frac{{\rm d}W_{12}(r,T)}{{\rm d}r}\;.
\label{alp_qq}
\end{eqnarray}
The coupling
$\alpha_{\rm qq}(r,T)$ then exhibits a maximum for fixed temperature at a certain distance denoted by
$r_{\rm max}$. By analyzing  the size of the maximum at $r_{\rm max}$
 we fix the temperature dependent coupling,
$\alpha_{\rm V}(T)$, as
\begin{eqnarray}
\alpha_{\rm V}(T)&\equiv&\alpha_{\rm qq}(r_{\rm max},T)\;.\label{alp_Tdef}
\end{eqnarray}
Before we start with the calculation of $\eta/s$ two important aspects have to be pointed out: first we
only consider the contribution of our dynamical degrees of freedom (quarks and antiquarks) for $\eta/s$,
since here the  gluons are massless and interaction free with respect to each other.
The fermion-gluon interaction is only space-like and
included in the potential (\ref{pot}).
Therefore, the gluons do not contribute to the ratio $\eta/s$, in contrast to
QCD perturbation theory, where the contribution of the quarks and of the
gluons are roughly of the same order~\cite{Arnold:2000dr,Arnold:2003zc}.
We thus consider the quark specific viscosity, i.e.
\be\label{eqn:Visk-qq}
\frac{\eta}{s_{\rm q}}=\frac{4}{5}\frac{T}{s_{\rm q}\gs_{\rm t}},
\ee
where the quark contribution of the entropy density $s_{\rm q}$ is
\be\label{eqn:s_q}
s_{\rm q}=s-s_{\rm g},
\ee
with the gluon contribution to the entropy density in the SB limit given by
\be\label{eqn:s_g}
s_{\rm g}=\frac{32\pi^2}{45}T^3.
\ee
The second important aspect is the applicability of Eq. (\ref{Eq:transport_xs}) to calculate
the transport cross section. We recall that
this expression for $\gs_{\rm t}$ has been derived for a Debye screened
interaction~\cite{Zhang:1999rs}, whereas the effective potential $W_{12}$
contains not only such a term  but also a purely screened part $\sim {\mathcal C}_2$.
A preliminary analysis, which actually has lead to the final formulation of Ref.~\cite{Mattiello:2009fk},
shows that the main contribution to the virial coefficient is given by the Coulomb screened part
of the potential. This confirms the observations of Ref.~\cite{Megias:2007pq}, that the exact value
of ${\mathcal C}_2$, which modulates the purely screened part of the potential, is not so important
for a good reproduction of the free energy.
Accordingly,  we may use Eq. (\ref{Eq:transport_xs}) as a good approximation
for the transport cross section.

The advantage of this formalism is that for the Debye screening mass no
further approximations must be done, as, for example, the assumtion of the
validity of the perturbation theory that leads to the specific expression given in
Eq. (\ref{def:screen-mass}) and is used often in the literature~\cite{Hirano:2005wx,Thoma:2004sp}.
In our model, the mass is directly given by Eq.(\ref{def:Debyemass}) and in this way
is calculated independently of the coupling $\ga_{\rm V}$ but,
at the same time, consistently -in a thermodynamic sense- with the properties of the QGP
because the parameter $\tilde g=1.3$ is fixed by thermodynamic quantities.
This demonstrates the important difference to other effective approaches used to understand
the enhancement of the parton transport cross section, that was concluded from the elliptic
flow measurements at RHIC.

We, furthermore, point out an analogy between strong
electromagnetic and strong quark-gluon plasma coupling presented
first in Ref.~\cite{Thoma:2004sp}: The Debye screening length,
i.e. $r_{\rm D}=\mu_{\rm scr}^{-1}$, cannot be used as a cut-off parameter
for the calculation of $\gs_{\rm t}$ because the interaction range
is larger than $r_{\rm D}$. Studies for complex non-relativistic
electromagnetic plasmas show the importance of these effects as
demonstrated in Ref.~\cite{PhysRevE.66.046414}. Furthermore, by
assuming a non-screened Coulomb interaction a modified Coulomb
logarithm $\gL^*$ was derived, which leads to a sizeable
enhancement of the transport cross section. The heuristic
translation of these results for the sQGP (given
in~\cite{Thoma:2004sp}) requires the screening mass in the
transport cross section to be replaced by \be\label{eqn:replace}
\mu_{\rm scr}\longrightarrow\frac{\mu_{\rm scr}}{4.6}\simeq 1.3 T \simeq \tilde{g} T.
\ee Surprisingly this is -quantitatively- the same temperature
dependence of the Debye mass as that obtained within our formalism
(cf. Eq. (\ref{def:Debyemass}) and discussion below).

Some additional comments are in place: Whereas in our virial
expansion approach the screening mass is calculated from the
thermodynamic quantities and the formalism includes crucial
ingredients such as relativity and an effective screened potential
- retaining thermodynamical  consistency -  the results of
Ref.~\cite{Thoma:2004sp} are obtained within several
approximations, i.e. a non-relativistic treatment, an unscreened
electromagnetic interaction and (though motivated but) {\em ad
hoc} modifications of the transport cross section. In this sense
the latter approach is heuristic: the modifications do not
automatically follow from the formalism used but have been
inserted {\em ad hoc} to achieve a better description of the
experimentally findings.

\section{Results}\label{Sec:Results}

\begin{figure}[ht]
\includegraphics[width=0.5\textwidth]{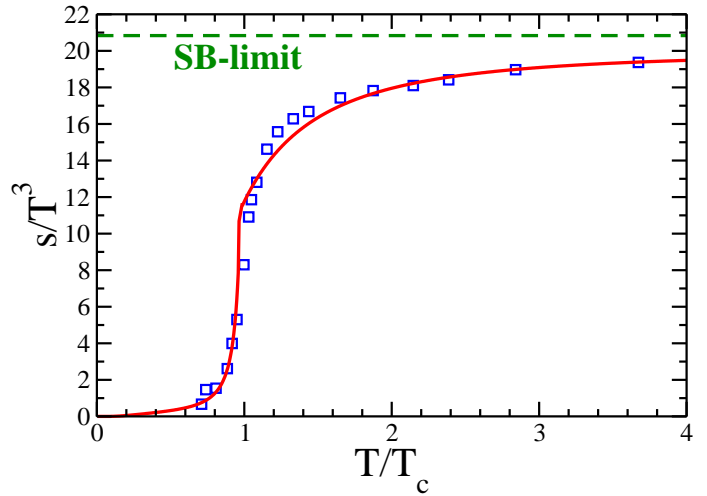}
\caption{\label{fig:S}
(Color online) Entropy density $s$ of the QGP as a function of the temperature
divided by $T^3$ from the virial expansion (solid line) using $\tilde g=1.30$. For
comparison the corresponding SB limit is displayed by the dashed line. The lQCD results
(open squares) have been adopted from Ref.~\cite{Cheng:2007jq}. }
\end{figure}

Before we discuss the results for $\eta/s$ from the virial
expansion approach, we demonstrate the applicability of our
formalism for the entropy density in comparison to the lQCD
calculations from Ref.~\cite{Cheng:2007jq}. In Fig~\ref{fig:S} the
entropy density $s$ (divided by $T^3$) is shown as a function of
the temperature expressed in units of the critical temperature
$T_{\rm c}$ from the virial expansion approach using Eq.
(\ref{eqn:Visk-s}) (solid line) as well as in the SB limit (dashed
line). The symbols denote the lQCD calculations from Ref.~\cite{Cheng:2007jq}.
For completeness we show also the entropy
density in the confined phase below $T_{\rm c}$, where we have
calculated all thermodynamic quantities (cf. Ref.~\cite{Mattiello:2009fk})
within a generalized resonance-gas model
and matched the different phases at equal pressure. Near
$T_{\rm c}$ the deviation of our results from the ideal gas limit
are sizeable and huge in the confined phase.
\begin{figure*}[ht]
\begin{center}
\includegraphics[width=0.6\textwidth]{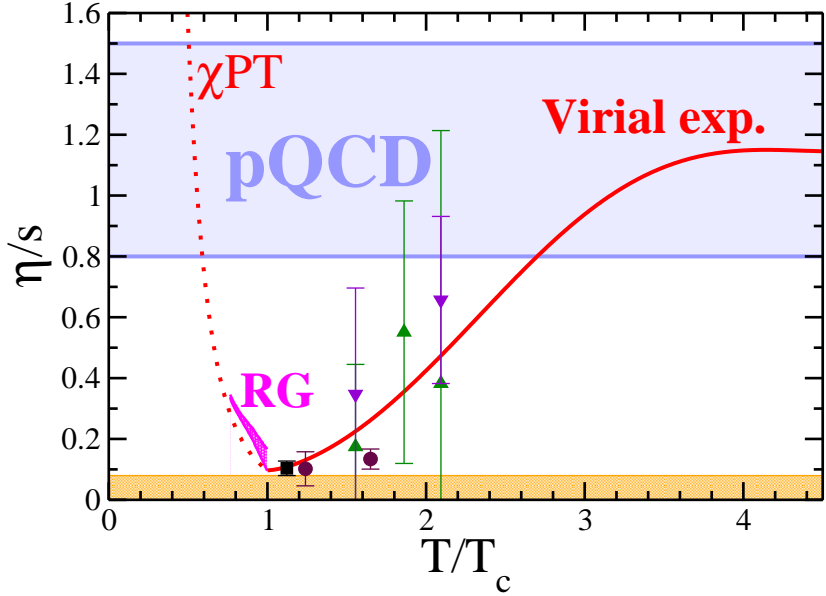}
\caption{\label{fig:VPL-EoS}
(Color online)  The  viscosity/entropy density ratio $\eta/s$
as a function of the temperature
expressed in units of the critical temperature $T_{\rm c}$ for $T/T_{\rm c}
<1$ and $T/T_{\rm c} > 1$. The lattice results are from
Ref.~\cite{Nakamura:2004sy} (triangles) and from
Ref.~\cite{Meyer:2007ic} (full dots). The different lines denoted by $\chi{\rm
  PT}$, ${\rm RG}$ and ${\rm Virial}$ ${\rm exp.}$ stand for the results from
the scaling behavior of the chiral perturbation
theory, the hadron resonance gas and the virial expansion approach (for
$T/T_{\rm c} \geq 1$), respectively (see text).}
\end{center}
\end{figure*}

The ratio of the viscosity to the entropy density has been calculated using
Eq. (\ref{eqn:Visk-qq}), where the quark contribution of the entropy density
$s_{\rm q}$ is given by Eqs. (\ref{eqn:s_q}) and (\ref{eqn:s_g}).
For the transport cross section the general expression given
in Eq. (\ref{Eq:transport_xs}) has been implemented,
where the effective temperature dependent coupling $\alpha_{\rm V}(T)$ is given in
Eq. (\ref{alp_Tdef}) and the Debye screening mass is  given by
Eq. (\ref{def:Debyemass}).
In Fig.~\ref{fig:VPL-EoS} our main results are presented in
comparison to other estimates. In the deconfined region,
$T/T_{\rm c}\geq1$, the solid red line show the results for
$\eta/s$ as a function of the temperature (in  units of the
critical temperature $T_{\rm c}$) using the EoS (\ref{eqn:s_q}-\ref{eqn:s_g}), the coupling
$\alpha_{\rm V}$ (\ref{alp_Tdef}) and the Debye mass (\ref{def:Debyemass})
derived within the virial expansion approach. Additionally, the experimental
point (square) from~\cite{Lacey:2009ps} and the lattice data from
Ref.~\cite{Nakamura:2004sy} (triangles) and from
Ref.~\cite{Meyer:2007ic} (full dots) are shown for comparison. In
the confined phase, $T/T_{\rm c}<1$, the purple region close
to $T_{\rm c}$ shows the estimates for $\eta/s$ in the resonance-gas
model~\cite{NoronhaHostler:2008ju}, including all the known
particles and resonances with masses $m<2$ GeV and also an
exponentially rising level density of Hagedorn states for $m>2$
GeV.
The dotted (red) line shows the scaling $\eta/s\propto T^{-4}$
combined with the requirement that $\eta/s=1/4\pi$ at $T_{\rm c}$.
This scaling behavior at low temperature has be found within
chiral perturbation theory in Ref.~\cite{Chen:2006iga}.
Additionally, the range $(0.8-1.5)$ for $\eta/s$ from perturbative
QCD (pQCD) from Ref.~\cite{Nakamura:2004sy} is sketched as a blue
region. Furthermore, the KSS lowest bound is also indicated by the
orange area.

In the deconfined phase several features become evident:
\begin{itemize}
\item[{\em i)}]  At $T_{\rm c}$ our result for $\eta/s\approx 0.097$ is very close to the theoretical
bound of $1/(4 \pi)$.
Additionally, the convexity of the specific viscosity near $T_{\rm c}$  suggests
a minimum close to $T_{\rm c}$ as expected in Refs.~\cite{Lacey:2006bc,Csernai:2006zz}.
\item[{\em ii)}] An almost linear increase of $\eta/s$ with the temperature is found
for $1.5T_{\rm c}\lesssim T\lesssim 3T_{\rm c}$.
\item[{\em iii)}] Currently the different lattice calculations are unable to
  provide quantitatively reliable results; the large error bars of the lattice data do not
allow for a conclusive comparison. Qualitatively, the increasing
behavior of the specific viscosity with the temperature might be
confirmed. In contrast, the experimental point (full square close
to $T_{\rm c}$) is reproduced very well by our result within the virial
approach.
\item[{\em iv)}] At higher temperatures a saturation of the ratio $\eta/s$
is found which roughly coincides within
the band for pQCD.
\end{itemize}
This last finding is supported by the observation that the entropy density $s$ as well as the
viscosity scale as $\propto T^{-3}$ for high temperatures. For the
entropy density $s$ this is in line with the SB limit,
which is approximately reproduced by lattice~\cite{Cheng:2007jq} and model
calculations~\cite{Cassing:2007yg,Mattiello:2009fk} at high
temperature.
For the shear viscosity $\eta$ also very different approaches show this functional dependence:
the strong quark-gluon plasma from AdS/CFT~\cite{Gubser:1998nz}, the quasiparticle
approximation with differently modeled quark selfenergy~\cite{Alberico:2007fu,Czerski:2007ns}
as well as the weak coupling estimate from Ref.~\cite{Thoma:1991em}.

With respect to the confined phase the resonance-gas as well as the $\protect\chi$PT calculation
suggest a decreasing behavior of $\eta/s$ with increasing temperature.
The 'constrained' scaling behavior ($\sim T^{-4}$)
slightly underestimates the reson\-ance-gas
calculation from Ref.~\cite{NoronhaHostler:2008ju}. Note, however, that the
entropy density in the Hagedorn
resonance model sensitively depends on the level density employed for
the continuum states. Accordingly, we show these results in
Fig.~\ref{fig:VPL-EoS} by a shaded band.
Choosing the scaling function ($\sim T^{-4}$) for the confined phase and our virial calculation for
the deconfined phase we obtain a well connected description for $\eta/s$ in the whole
temperature range (up to 5 $T_{\rm c}$) which clearly shows a minimum near $T_{\rm c}$ close to the KSS bound.

To investigate in more detail the possibility of a minimum of $\eta/s$ at the critical
temperature from the virial expansion calculation we search for an approximation
of the specific viscosity in the deconfined phase around $T_{\rm c}$.
We employ a Taylor expansion of $\eta/s$ as a function of $t\equiv T/T_{\rm c}$ at $t=1$, i.e.
\be\label{eqn:Taylor}
\frac{\eta}{s}=\sum_{n=0}^{\infty}a_n(t-1)^n=\sum_{n=0}^{N}a_n(t-1)^n+R_N,
\ee with \be\label{eqn:coeff}
a_n=\frac{1}{n!}\frac{{\rm d}^n(\eta/s)}{{\rm d}^nt}(t=1).
\ee
In (\ref{eqn:Taylor}) $N$ indicates the order of the Taylor polynomial while $R_N$ stands
for the rest of the corresponding expansion. Our stra\-tegy is now
to investigate how different polynomial approximations - labeled
by different $N$ - can describe $\eta/s$ up to $T=2T_{\rm c}$,
$T=3T_{\rm c}$ and $T=4.5T_{\rm c}$. For each $N$ we calculate the
$\chi^2/{\rm dof}$ to evaluate the quality of the approximation.
Since the expansion coefficients are connected to derivatives of
the specific viscosity at $T_{\rm c}$ (see Eq.(\ref{eqn:coeff}))
by requiring that some coefficients vanish, we automatically impose
that the corresponding derivatives are equal to zero. In
particular, with the requirement $a_1=0$, we may investigate the
possibility of a minimum at the critical temperature.
Some polynomial fits to our results are given in the
Tables~\ref{tab1},~\ref{tab2} and~\ref{tab3}.

\begin{table}
\caption{Results for the fit of $\eta/s$ up to $T=2T_{\rm c}$.}
\centering
\begin{tabular}{ccccc}
\hline\hline
N & $a_1$ & $a_2$&$a_3$&$\chi^2/{\rm dof}$\\ [0.5ex]
\hline
2& 0.0968&0.2359& -&0.00102\\
2& 0&0.3566& -&0.03959\\
3& 0.0616&0.3528& -0.08747&5.1 $10^{-5}$\\
3& 0&0.5373& -0.2163&0.00255\\[1ex]
\hline
\end{tabular}
\label{tab1}
\end{table}

\begin{table}[t]
\caption{Results for the fits of $\eta/s$ up to $T=3T_{\rm c}$.}
\centering
\begin{tabular}{ccccc}
\hline\hline
N & $a_2$&$a_3$&$a_4$&$\chi^2/{\rm dof}$\\[0.5ex]
\hline
2& 0.2477&-& -&1.4478\\
3& 0.4469&-0.1194& -&0.0164\\
4& 0.5044&-0.1998& 0.0268&0.0061\\[1ex]
\hline
\end{tabular}
\label{tab2}
\end{table}
As expected, the power $N=2$ cannot describe the behavior of $\eta/s$ for $T \ge 2T_{\rm c}$.
However, the inclusion of higher power allows to reproduce very well our data also for higher
temperatures.
For $N=3$ and $N=4$ we find a very good approximation for our calculated
results for $\eta/s$ up to $T=3T_{\rm c}$ and  $T=4.5T_{\rm c}$, respectively.
Therefore, a power law approximation with a vanising or very small linear coefficient suggests a
simple parametrization for the specific viscosity which includes a minimum
close to $T_{\rm c}$.
\begin{table}[h]
\caption{Results for the fit of $\eta/s$ up to $T=4.5T_{\rm c}$.}
\centering
\begin{tabular}{ccccc}
\hline\hline
N & $a_2$&$a_3$&$a_4$&$\chi^2/{\rm dof}$\\[0.5ex]
\hline
2& 0.1227&-& -&43.5251\\
3& 0.3756&-0.0845& -&0.8524\\
4& 0.4955&-0.1781& 0.01738&0.0228\\[1ex]
\hline
\end{tabular}
\label{tab3}
\end{table}

\section{Conclusions}\label{Sec:Concl}
We have performed an investigation of the specific viscosity
$\eta/s$ in the QGP in a dynamical way within kinetic theory using the virial expansion
approach introduced in Ref.~\cite{Mattiello:2009fk}. In this
context the investigation of the interaction between the partons
in the deconfined phase plays a crucial role to reproduce the
thermodynamic properties of the QGP in comparison to lattice QCD.
By using a generalized classical virial expansion we have
calculated the corrections to a single-particle partition function
starting from an interaction potential whose parameters are fixed
by thermodynamical quantities. Furthermore, in the virial
expansion approach we can directly extract the coupling $\ga_{\rm
V}$ to be employed for the determination of the transport cross
section which enters the ratio $\eta/s$. We find $\eta/s\approx 0.097$ at
$T_{\rm c}$ which is very close to the theoretical lower bound.
Furthermore, for $T\leq 1.8 T_{\rm c}$  the ratio is in the range
of the experimental estimates $0.1-0.3$ extracted from RHIC
experiments.

Additionally, a detailed analysis of the temperature dependence of
our results for $\eta/s$ has been performed. Within a Taylor
expansion around the critical temperature we found that a power
law with a vanising or very small linear coefficient suggests a simple
parametrization for the specific viscosity. This indicates the
existence of a minimum in $\eta/s$ close to $T_{\rm c}$.

Since we focus on the deconfined phase only, we do not
investigate whether in the vicinity of $T_{\rm c}$ a phase transition or a rapid
cross over occurs~\cite{Khvorostukhin:2010aj}. However, our approach, which provides a unified description
of the QGP thermodynamic as well as of its sheer viscosity, is a first
important improvement in the description of the quark-gluon plasma beyond
mean-field models~\cite{Csernai:2006zz}.

For further work, using relativistic molecular dynamical simulations, where
the phenomenological quark-quark interaction used here can be
implemented as well as the QCD equation of state,
we may calculate further correlations in the partonic phase by considering partons in a box
with periodic boundary conditions at fixed energy density (or temperature).
Furthermore, such molecular dynamical calculations will allow to study partonic systems
also out of equilibrium and provide important insight on the dynamics of hadronization.

\vspace{1cm}
{\em Acknowledgements:} Work supported in part by the Deut\-sche
Forschungsgemeinschaft (DFG).




\bibliographystyle{epjc}
\bibliography{literatureEPJC}

\end{document}